\documentclass[aps,prl,showpacs,twocolumn]{revtex4}
\usepackage{graphicx}
\usepackage{amsmath}
\begin{document}
\title{The effect of partonic wind on charm quark correlations in
high-energy nuclear collisions}
\date{\today}
\author{X.~Zhu$^{1}$, N.~Xu$^{2}$ and P.~Zhuang$^{1}$}
\affiliation{$^1$Physics Department, Tsinghua University, Beijing
100084, China}
\affiliation{$^2$Nuclear Science Division, Lawrence Berkeley
National Laboratory, Berkeley, CA 94720, USA}

\begin{abstract}
In high-energy collisions, massive heavy quarks are produced
back-to-back initially and they are sensitive to early dynamical
conditions. The strong collective partonic wind from the fast
expanding quark-gluon plasma created in high-energy nuclear
collisions modifies the correlation pattern significantly. As a
result, the angular correlation function for D$\overline{\rm D}$
pairs is suppressed at the angle $\Delta\phi=\pi$. While the hot
and dense medium in collisions at RHIC ($\sqrt{s_{NN}}=200$ GeV)
can only smear the initial back-to-back D$\overline {\rm D}$
correlation, a clear and strong near side D$\overline{\rm D}$
correlation is expected at LHC ($\sqrt{s_{NN}}=5500$ GeV).
\end{abstract}
\maketitle

The ultimate goal of relativistic heavy ion collision experiments
at RHIC and LHC is to detect the formation of a possible new state
of matter, the so-called quark-gluon plasma (QGP), created in the
early stage of the collisions. An essential difference between
elementary particle collisions and nuclear collisions is the
development of the collective motion in the latter. The partonic
collectivity is one of the consequencies of the formation of QGP.
The collective flow of all hadrons, especially the multi-strange
hadrons $\phi$ and $\Omega$, has been experimentally
measured~\cite{star1,star2,star3,star31}. These results are
consistent with the predictions based on hydrodynamic model
calculations~\cite{sorge1,sorge2,ollitrault,teaney,huovinen} and
demonstrate strong collective expansion at RHIC. However, the
issue of local thermalization of the partonic matter remains to be
addressed.

The observables related to heavy quarks are of particular interest in
the study of thermalization. Unlike light quarks which are controlled
by chiral dynamics and are almost massless in the deconfined phase
where the chiral symmetry is restored, heavy quarks remain massive in
QGP and can only be pairly-created in the early stage of the
collisions at RHIC. In addition, during the subsequent evolution, the
number of heavy quarks is unchanged, because the typical achievable
temperature of the medium is in general lower than the thresholds for
heavy quark production.  However, the heavy quarks can participate in
collective motion if, and only if, their interactions at partonic
level occur at a sufficiently high frequency.  As the light quark and
gluon densities are much higher than that of the heavy quarks, any
collective motion of heavy quark hadrons will be a useful tool for
studying the early thermalization of light quarks in high-energy
nuclear collisions.

From lattice calculations~\cite{karschQM06}, the thermodynamic
functions can not reach their Stefan-Boltzmann limits even at a
temperature of four times the critical value, indicating that the
system is still strongly coupled at very high temperature.
Recently, the elliptic flow and nuclear modification factors of
heavy quarks' decay electrons are measured~\cite{star32,phenix1}
at RHIC. In order to explain the data, a large drag coefficient in
the Langevin equation, describing the propagation of charm quarks
in the strongly coupled QGP, is necessary~\cite{hees,moore}. Two-
and three-body interactions~\cite{wicks,liu} of heavy quarks seem
to become important as well. Furthermore, from the study of
$J/\psi$ regeneration in QGP~\cite{yan}, we have found that
thermalized charm quarks can better describe the $J/\psi$ averaged
transverse momentum square~\cite{phenix2,yan}. All these analysis
suggest that heavy quarks have been participating in frequent
rescatterings at the early partonic stage in high-energy nuclear
collisions.

In hard scatterings, heavy quark pairs are produced back-to-back
at leading order. A typical example for the processes is the gluon
fusion: $gg \to c\overline{\rm c}$. In hadron-hadron collisions,
although the initial $k_T$ kick, high order processes and
fragmentation smear the angular correlations, the back-to-back
peak has been observed in the final state D$\overline{\rm D}$
correlation~\cite{lourenco}. In high-energy nuclear collisions, on
the other hand, a hot and dense partonic medium is expected at the
early stage. The interactions between the charm quarks and the
medium will modify the angular correlations and even lead to a
complete absence of this back-to-back correlation~\cite{zhu1} in
the final D-meson correlation function.

As we emphasized above, the QGP formed in the early stage of
high-energy nuclear collisions is not a static system, but a fast
expanding fireball. The heavy quarks, which frequently interact
with QGP medium, will experience the strong collective flow $-$
the partonic wind. The effect of the partonic wind on the
D$\overline{\rm D}$ correlation function can be estimated from the
transformation between two reference frames. Suppose the c and
$\overline{\rm c}$ in a pair move radially in QGP and the velocity
of the one which propagates towards the center of the fireball is
$v_{\rm c}$ in the rest frame of the fluid element. When $v_{\rm
c}$ is less than the radial flow $v_f$ of the fluid element with
respect to the laboratory frame, both charm quarks will move
together towards outside and the initial back-to-back correlation
is turned to the same side by the partonic wind! In central Au+Au
collisions at RHIC, taking $v_f \simeq 0.6 c$~\cite{star4}, the
critical charm quark momentum for the turning of the
c$\overline{\rm c}$ correlation is in the order of 1 GeV. Taking
into account both the partonic wind and the scatterings between
the heavy quarks and the medium, the turning could occur at a much
higher momentum. As a result, the partonic wind will speed up the
disappearance of the D$\overline{\rm D}$ correlation and may even
change the back-to-back correlation to the near side when the wind
is strong enough.

In this Letter, we focus on the mid-rapidity D$\overline{\rm D}$
azimuth correlation in p+p and A+A collisions at RHIC and LHC
energies.  We first estimate the charm hadron correlation for p+p
collisions with PYTHIA~\cite{pythia}. The motion of charm quarks
in the medium is treated as a random walk by the Langevin
equation~\cite{svetitsky1}. The medium itself is described by a
2+1 dimensional Bjorken type hydrodynamic
calculation~\cite{zhu2,yan} with proper initial conditions at
corresponding collision energies. The effect of the partonic wind
on the final D$\overline{\rm D}$ correlation is obtained by
solving the coupled Langevin equation with the hydrodynamic
evolutions.

\begin{figure} [h]
\includegraphics[width=0.48\textwidth]{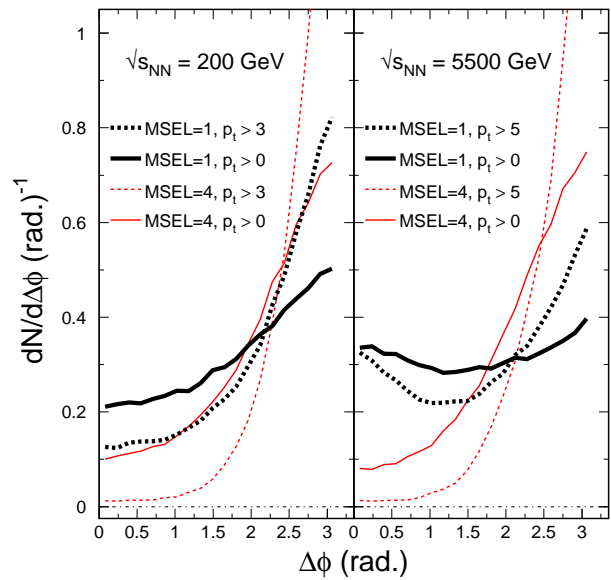}
\vspace{-0.75cm} \caption{(Color online) The D$\overline {\rm D}$
correlation as a function of relative azimuth angle $\Delta\phi$
in p+p collisions at $\sqrt{s} = 200$ GeV (left) and $\sqrt{s} =
5500$ GeV (right), calculated by PYTHIA (v. 6.327)~\cite{pythia}.
The transverse momentum depedence is shown by the $p_t$ cuts (GeV)
on single D-mesons. MSEL=4 (thin lines) and MSEL=1 (thicker lines)
indicate, respectively, the calculation to leading order and to
next-to-leading order.} \label{fig1}
\end{figure}

We start by reviewing the D$\overline{\rm D}$ correlation in pp
collisions. The Monte Carlo event generator PYTHIA~\cite{pythia}
reproduces well the experimentally observed correlation measured
at fixed target energies~\cite{lourenco}. Fig.\ref{fig1} shows the
calculated correlation for p+p collisions at RHIC and LHC
energies. Compared to the calculation~\cite{zhu1} at leading order
(LO) pQCD which contains only pair creation processes, the
next-to-leading order (NLO) pQCD calculation includes pair
creation with radiative corrections, flavor excitation and gluon
splitting by switching on the initial and final parton
showers~\cite{norrbin,carrer}. Although this treatment is not
exactly at next-to-leading order, some aspects of the
multiple-parton emission phenomenon are
reproduced~\cite{norrbin,carrer}. Since c$\overline{\rm c}$ pairs
from flavor excitation and gluon splitting do not show explicit
angular correlation, the NLO D$\overline{\rm D}$ correlation
becomes weak at RHIC energy and even vanishes at LHC energy,
compared to the corresponding LO result. At RHIC energy, the
correlation of D$\overline{\rm D}$ pairs with high transverse
momentum $p_t$ is mostly preserved, since they are mainly from the
pair production processes. However, at LHC energy, even for pairs
with $p_t$ larger than 5 GeV/c, the back-to-back correlation is
strongly suppressed, since gluon splitting and flavor excitation
become dominant at very high energy~\cite{carrer}.

%--============================== Figure 1

Since there is no obvious D$\overline {\rm D}$ correlation in p+p
collisions at LHC energy, the measurement of smearing this
correlation in heavy ion collisions becomes meaningless at LHC.
However, the medium temperature at LHC is so high that heavy
quarks are much easier to get thermalized with the medium, and
then flow with the medium and be pushed by the partonic wind to
the same direction. Therefore, near side, instead of back-to-back,
D$\overline{\rm D}$ correlation might be possible at LHC.

To explore whether the partonic wind generated in central
relativistic nucleus-nucleus collisions at RHIC and LHC influences
the D$\overline{\rm D}$ angular correlation, we employ a
non-relativistic Langevin approach which describes the random walk
of charm quarks in a QGP and was first described in
Ref.~\cite{svetitsky1},
\begin{equation}
\label{langevin}
d\vec{p}/dt =-\gamma(T)\vec{p}+\vec{\eta},
\end{equation}
where $\vec{\eta}$ is a Gaussian noise variable, normalized such
that $\langle \eta_i(t)\eta_j(t')
\rangle=\alpha(T)\delta_{ij}\delta(t-t')$ with $i,j$ indexing
directions. Both the drag coefficient $\gamma$ and the
momentum-space diffusion coefficient $\alpha$ depend on the local
temperature $T$. We use the same parameterization for $\gamma$ as
in Ref.~\cite{svetitsky1}, $\gamma(T) = a T^2$, with $a = 2\cdot
10^{-6}$ (fm/$c$)$^{-1}$ MeV$^{-2}$ and also neglect the momentum
dependence of $\gamma$. $\alpha$ can be calculated from the
fluctuation-dissipation relation in equilibrium, $\langle p_i^2
\rangle=\alpha/2\gamma$, with $\langle p_i^2\rangle=1.33\, m_{\rm
c}\, T$ and charm quark mass $m_{\rm
c}=1.5$~GeV/$c^2$~\cite{svetitsky1}.

The above drag coefficient is estimated with pQCD adopting a large
coupling constant $\alpha_s=0.6$~\cite{svetitsky2}. The recent
pQCD calculations~\cite{hees,golammustafa} show a factor of 2-3
smaller drag coefficient, but the quasi-hadronic resonant states
in QGP at moderate temperature would result in a much larger drag
coefficient~\cite{hees}. Since exact values of the drag
coefficient from lattice QCD are not yet available, we will choose
in the following numerical calculations the drag parameter $a$ in
a reasonable region. The comparison of a similar Langevin
calculations~\cite{moore} for $R_{{\rm AA}}$ and $v_2$ of
non-photonic electrons with recent PHENIX data shows that the data
favor a large drag coefficient~\cite{phenix1}. We checked that the
drag parameter $a$ corresponding to the diffusion coefficient
$D=3/(2\pi T)$ in~\cite{moore,horowitz} is about $6\cdot 10^{-6}$
(fm/$c$)$^{-1}$ MeV$^{-2}$.

For simplicity, we focus on charm quarks generated at mid-rapidity
in central collisions. We assume that the produced partonic plasma
reaches local equilibrium at time $\tau_0$. After that, the plasma
evolves according to the 2+1 dimensional Bjorken's hydrodynamics
which describes the collective flow of the bulk matter very well
at RHIC~\cite{kolb,zhu2},
\begin{eqnarray}
\label{hydro}
\partial_{\tau}E+\nabla\cdot{\bf M} &=& -(E+p)/{\tau}\
,\nonumber\\
\partial_{\tau}M_x+\nabla\cdot(M_x{\bf v}) &=& -M_x/{\tau}-\partial_xp\ ,
\nonumber\\
\partial_{\tau}M_y+\nabla\cdot(M_y{\bf v}) &=& -M_y/{\tau}-\partial_yp \
,\nonumber\\
\partial_{\tau}R+\nabla\cdot(R{\bf v}) &=& -R/{\tau}
\end{eqnarray}
with the definitions $E=(\epsilon+p){\tilde\gamma}^2-p$, ${\bf
M}=(\epsilon+p){\tilde\gamma}^2{\bf v}$ and $R=\tilde\gamma n$,
where $\tilde\gamma$ is the Lorentz factor, and $\epsilon, p$ and
${\bf v}$ are the energy density, pressure and transverse velocity
of QGP. To close the hydrodynamic equations, we take the equation
of state of ideal quark-gluon plasma~\cite{sollfrank} with first
order phase transition at $T_c = 165$ MeV. The initial condition
for central collisions of heavy nuclei at RHIC is the same as in
~\cite{zhu2}. For Pb+Pb collisions at LHC energy, the initial
baryon density is neglected and the maximum initial entropy
density is assumed to be 5.5 times the value (about 110 fm$^{-3}$
~\cite{kolb2}) at RHIC. Taking into account the earlier
equilibrium time at LHC, this initial value is reasonable
according to the multiplicity prediction at LHC~\cite{fini}. For a
simple comparison, the maximum temperature $T_0$ is 340 MeV at
$\tau_0=0.6$ fm/c at RHIC and 610 MeV at $\tau_0=0.3$ fm/c at LHC.

Before the time $\tau_0$, the plasma changes rapidly and is not
fully equilibrated. To simplify the calculation, we assume, as
done in~\cite{svetitsky1}, that the plasma is already in
equilibrium before $\tau_0$ with a constant temperature $T=T_0$,
and the charmed quarks diffuse in the same way as they do after
$\tau_0$.

The initial charm quarks are generated with PYTHIA at a time of
the order of $1/m_c\simeq$~0.1~fm/$c$. In order to isolate the
effects purely due to parton-parton rescattering in the outlined
medium, we use a delta function for fragmenting a charm quark into
a charmed hadron at the hadronization stage.

The radius $r$ where each pair is created is randomly generated
from a distribution reflecting the number of binary
nucleon-nucleon collisions taking place at that radius,
$p(r)\propto(R_A^2-r^2)2\pi r $, where $R_A$ is the geometric
radius of the colliding nuclei.

The evolution of the charm momentum with the Langevin equation
ends when $T$ reaches the critical temperature $T_{\rm c}$ or when
the charm quark leaves the QGP volume. To get a first estimation
of the QGP effect on the pair correlation, we omit the
contribution from the mixed phase and hadronic phase.

It is necessary to note that the hydrodynamic equations determine
not only the local temperature $T$ which characterizes the drag
coefficient but also the local fluid velocity ${\bf v}$ which
controls the partonic wind effect on the c$\overline{\rm c}$
correlation. Since the Langevin equation is defined in the local
rest frame of fluid element, we should make a transformation from
the rest frame to the laboratory frame which is governed by the
fluid velocity ${\bf v}$.

%--============================== Figure 2
\begin{figure} [h]
\includegraphics[width=0.48\textwidth]{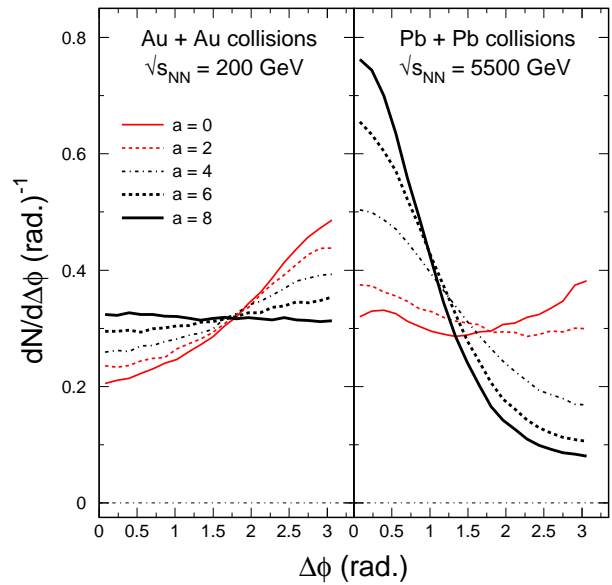}
\vspace{-0.75cm} \caption{(Color online) The D$\overline{\rm D}$
correlation as a function of relative azimuth angle $\Delta \phi$
with different drag parameter $a$ for central Au+Au collisions at
RHIC (left) and Pb+Pb collisions at LHC (right). The unit of the
drag coefficient $a$ is [10$^{-6}$(fm/c)$^{-1}$MeV$^{-2}$]. In the
PYTHIA calculation to next-to-leading order, no $p_t$ cut was
applied. The initial temperature $T_0$ and thermalization time
$\tau_0$ are, respectively, 340 MeV and 0.6 fm/c at RHIC and 610
MeV and 0.3 fm/c at LHC.} \label{fig2}
\end{figure}

Fig.~\ref{fig2} shows the results of D$\overline{\rm D}$ angular
correlation for central $\sqrt{s} = 200$A GeV Au+Au collisions at
RHIC (left plot) and $\sqrt{s} = 5500$A GeV Pb+Pb collisions at
LHC (right plot). The thin solid lines with $a=0$ are the
corresponding results directly from PYTHIA with parton showers
turned on. At RHIC, with increasing drag coefficient the initial
back-to-back D$\overline{\rm D}$ correlation gradually become more
and more flat.  At $a \sim 6\cdot10^{-6}$ (fm/c)$^{-1}$MeV$^{-2}$,
the correlation is totally washed out, see left plot. Note that
the recent PHENIX data~\cite{phenix1} seem to favor the relatively
large value of the drag coefficient. The experimental measurement
of this smearing effect in the heavy quark correlation, especially
for those pairs at high $p_t$, can shed a light on the
thermalization status of the medium at RHIC~\cite{zhu1}. At LHC,
one starts with an almost flat correlation, see the thin solid
line in the right plot. The near side D$\overline{\rm D}$
correlation becomes more and more visible with increasing the drag
coefficient. For $a \sim 4\cdot 10^{-6}$(fm/$c$)$^{-1}$MeV$^{-2}$,
the near side D$\overline{\rm D}$ correlation is already very
strong. The larger the drag coefficient means the stronger the
interactions. This result indicates that c$\overline{\rm c}$ pairs
are thermalized with the medium quickly and they are moving in the
same direction as the partonic wind.

Note that there are more than one pair of D$\overline{\rm D}$ produced
in high energy nuclear collisions. At mid-rapidity $|y|
\le 1$, the total number of charm pairs is in the order of 2-4 and
6-10 at RHIC and LHC, respectively. The multiple production of
un-correlated D and $\overline{\rm D}$ mesons will give an
additional background to the measured angular correlations. Our
tests show that this background can be removed by the mixed event
method. The mixed-event subtracted correlation function should be
compared with the calculated results shown in Fig.2.

In summary, we investigated the partonic wind effect on
D$\overline{\rm D}$ angular correlation in high-energy nuclear
collisions. At RHIC, while there is still sizable D$\overline{\rm D}$
back-to-back correlation in p+p collisions, it is washed out in the
most central Au+Au collisions. At LHC, the pQCD high order
contributions to c-quark production become dominant and there is no
more clear D$\overline {\rm D}$ angular correlation even in elementary
collisions. However, charm quarks are thermalized with the medium
quickly at LHC. As a result, a near side D$\overline{\rm D}$
correlation, driven by the strong partonic wind, will replace the flat
angular distributions. Such dramatic changes in the D$\overline{\rm
D}$ angular correlations, originated from the strong interaction
between charm quarks and the hot medium, reflected by the large drag
coefficient ($\ge4\cdot 10^{-6}$(fm/$c$)$^{-1}$MeV$^{-2}$) in the
Langevin equation in our treatment, should be considered as a
signature of the strongly coupled QGP formation at RHIC and LHC.

{\it Acknowledgements} We thank Drs. M. Bleicher, H. St\"{o}cker
and K. Schweda for exciting discussions.  The work is supported by
the NSFC Grant No.10428510, the 973 project No.2007CB815000 and
the U.S. Department of Energy under Contract No.
DE-AC03-76SF00098.

%--==================================================================

\vfill\eject
\end{document}